\documentstyle[11pt]{article}
\begin{document}
\title{{\normalsize{{\hskip 9cm} KL-TH/95/21}} \\
2D quantum dilaton gravitational Hamiltonian, boundary terms and new
       definition for total energy}
\author{{H.J.W.M$\ddot{u}$ller-Kirsten${}^a$, Jian-Ge Zhou${}^a$, Yan-Gang
         Miao${}^b$ and J.-Q.Liang${}^c$}\\
{\small ${}^a$ Department of Physics, University of Kaiserslautern, P.O.Box
        3049,}\\
{\small D-67653 Kaiserslautern, Germany}\\
{\small ${}^b$ Department of Physics, Xiamen University, Xiamen 361005,
        China}\\
{\small ${}^c$ Institute of Theoretical Physics, Shanxi University, Taiyuan,
        Shanxi 030006, China}}
\date{}
\maketitle
\vskip 48pt
\begin{center}{\bf Abstract}\end{center}
\baselineskip 22pt
\par
    The ADM and Bondi mass for the RST model have been first discussed from
Hawking and Horowitz's argument. Since there is a nonlocal term in the RST
model, the RST lagrangian has to be localized so that Hawking and Horowitz's
proposal can be carried out. Expressing the localized RST action in terms of
the ADM formulation, the RST Hamiltonian can be derived, meanwhile keeping
track of all boundary terms. Then the total boundary terms can be taken as the
total energy for the RST model. Our result shows that the previous expression
for the ADM and Bondi mass actually needs to be modified at quantum level, but
at classical level, our mass formula can be reduced to that given by Bilal and
Kogan [5] and de Alwis [6]. It has been found that there is a new contribution
to the ADM and Bondi mass from the RST boundary due to the existence of the
hidden dynamical field. The ADM and Bondi mass with and without the RST
boundary for the static and dynamical solutions have been discussed
respectively in detail, and some new properties have been found. The thunderpop
of the RST model has also been encountered in our new Bondi mass formula.
\newpage
\baselineskip 22pt
    In recent years, the definition for the total energy in 2D dilaton
gravity has attracted a lot of attention [1-7]. The formula in refs.[1-4] for
the Arnowitt-Deser-Misner (ADM) mass of 2D dilaton gravity was found to be
incomplete, and the origin of this incompleteness can be traced to the implicit
assumption that ${\delta}{\phi}$ and ${\delta}{\rho}$ are
$O(e^{-2{\lambda}{\sigma}})$, but they actually contain terms
$O(e^{-{\lambda}{\sigma}})$ [5,6]. Bilal and Kogan [5] gave an improved
expression for ADM mass by imposing some asymptotic conditions, and assumed no
contribution to the ADM mass from ${\sigma}=-\infty$, but their mass
formula did not contain the quantum corrections. Later, de Alwis [6] obtained
the quantum corrected expressions for the ADM and Bondi mass
using arguments given
by Regge and Teitelboim [8], and found that there was a contribution to the ADM
and Bondi mass from the point ${\sigma}{\rightarrow}-{\infty}$. However, the
quantum versions of the original Callan-Giddings-Harvey-Strominger (CGHS) model
[1], for instance, the Russo-Susskind-Thorlacious (RST) model [9], have been
shown that there exists a hidden dynamical field [10-12], which was omitted
in
previous considerations of the semiclassical approach. So one may ask whether
there is a contribution to the total energy from this hidden dynamical field.

    On the other hand, in the above derivations of the 2D dilaton gravitational
Hamiltonian, the boundary term has been ignored. This results in a Hamiltonian
which is just a multiple of a constraint. Then one must add to this constraint
appropriate boundary terms so that its variation is well defined [5,6].
Recently, Hawking and Horowitz [13] proposed to keep track of all surface
terms in a general derivation of
the gravitational Hamiltonian starting from the Einstein-Hilbert action.
The resulted surface terms can
be taken as the definition of the total energy even for spacetime that is not
asymptotically flat [14].
Thus the boundary terms in H come directly from the boundary terms in the
action, and do not need to be added "by hand".  However, they just considered
the case of the 4D Einstein-Hilbert action, in adopting 2D quantum dilaton
gravity as a model for 4D gravity, it is important to know what features the
two theories have in common.  For example, is there a lower bound to the total
energy for the RST model?  and what role does the hidden dynamical field play
in the total energy expression?

   In the present paper, the ADM and Bondi mass
for the RST model are first discussed using the argument given by Hawking and
Horowitz [13].  Since there is a nonlocal term in the RST model, the RST
lagrangian must first be localized so that Hawking and Horowitz's proposal can
be carried out.  For this purpose, the scalar field $\chi$ and the boundary
term are introduced in order that the reformulated RST action is well-posed
and local [10,11,15].  Expresssing the RST action in terms of the ADM
formulation [15-19], the RST Hamiltonian can be derived, meanwhile keeping
track of all boundary terms.  Then the total boundary terms can be identified
as the ADM or Bondi mass for the RST model.  The result shows that the mass
formula used in refs.[1-7] actually needs to be modified at quantum level.
But at classical level, our mass formula can be reduced to that of refs.[5,6].
It has been shown that there is a new contribution to the total energy from
the hidden dynamical field.  In the absence of the RST boundary, that is, the
left boundary of the spacetime is ${\bar{\sigma}=-{\infty}}$, the ADM mass for
the static solutions is zero, whereas for the dynamical case with collapsing
matter, it has been found that there is an infinite contribution to the ADM
mass from the negative infinite end of the space, i.e., this theory describes
black hole collapse in an infinite bath of radiation.  The Bondi mass has also
been discussed.  However, in the case without the RST boundary, the Bondi mass
is found to be finite.  In the presence of the RST boundary, it has been shown
that the ADM mass for the static solutions with $m_0>0$ ($m_0$ will be defined
below) is zero, so we resolve the problem that was left in ref.[6].  In the
dynamical case, a new contribution to the ADM mass from the RST boundary has
been found, which is just the consequence that the hidden dynamical field
affects the mass formula.  When ${\bar{\tau}{\rightarrow}-{\infty}}$, the ADM
mass goes to the mass of the collapsing matter.  However, at an intermediate
time the Bondi mass becomes negative, and is discontinuous across a certain
null line, that is, the thunderpop of RST [9] has been encountered.  In the
region ${\bar{\sigma}}^{-} > {\bar{\sigma}}^{-}_{s}$
(${\bar{\sigma}}^{+}_{s}, {\bar{\sigma}}^{-}_{s}$
are the points where the apparent horizon and the critical curve
intersect), the black hole has decayed and the solution is taken to be linear
dilaton vacuum (LDV), but the Bondi mass is found to be nonzero due to the
existence of the hidden dynamical field, which reflects the fact that the
spacetime is not globally flat.  Our result also shows that with the new
definition for the total energy, the ADM and Bondi mass have a lower bound in
the presence of the RST boundary.

   We now consider the RST model with the action [9]
\begin{eqnarray}
S&=&\frac{1}{4\pi}\int
d^{2}{\sigma}\sqrt{-g}\left\{e^{-2\phi}\left[R+4({\nabla}{\phi})^{2}
+4{\lambda}^2\right]-{\frac 1 2}\sum_{i=1}^{N}({\nabla}f_{i})^2\right.
\nonumber \\
& &\left.-\frac{\kappa}{4}\left(R\frac{1}{{\nabla}^2}R+2{\phi}R\right)\right\}
\end{eqnarray}
where $g_{{\alpha}{\beta}}$ is the metric on the 2D manifold M, R is its
curvature scalar, $\phi$ is
the dilaton field, and the $f_i, i=1,\cdots,N$, are {\em N} scalar matter
fields.

  According to refs.[10,11,15], one can introduce an independent scalar field
$\chi$ to localize the conformal anomaly term, and add a boundary term to
define the variational problem properly. Then Eq.(1) turns into [10-11]
\begin{eqnarray}
S&=&\frac{1}{4\pi}\int d^{2}{\sigma}\sqrt{-g}\left\{R\tilde{\chi}+4[({\nabla}
  {\phi})^{2}+{\lambda}^2]e^{-2\phi}-\frac{\kappa}{4}g^{{\alpha}{\beta}}
  {\partial}_{\alpha}{\chi}{\partial}_{\beta}{\chi}\right.\nonumber \\
  & &\left.-{\frac 1 2}\sum_{i=1}^{N}({\nabla}f_{i})^2\right\}-\frac{1}{2\pi}
     \int d{\Sigma}\sqrt{{\pm}h}K\tilde{\chi}
\end{eqnarray}
where $\tilde{\chi}=e^{-2\phi}-\frac{\kappa}{2}(\phi-\chi)$, {\em h} is the
induced metric on the boundary of M, and K is
the extrinsic curvature of ${\partial}M$. Following the ADM formulation, the
metric can be parametrized as follows [15-19]:
\begin{equation}
g_{{\alpha}{\beta}}=e^{2\rho}{\hat{g}}_{{\alpha}{\beta}}
\end{equation}
\begin{equation}
{\hat{g}}_{{\alpha}{\beta}}=\left(
\begin{array}{cc}
   -{\omega}^2+{\theta}^2 & {\theta} \\
    {\theta}              &   1
\end{array}\right)
\end{equation}
where ${\omega}(x)$ and ${\theta}(x)$ are lapse and shift functions
respectively, and the conformal factor $e^{2{\rho}}$ has been factored out.

    In terms of this parametrization, the action (2) can be written as
\begin{eqnarray}
S&=&\frac{1}{2}\int d^{2}{\sigma}\sqrt{\hat{g}}\left\{\hat{R}\tilde{\chi}
   +2{\hat{g}}^{{\alpha}{\beta}}{\partial}_{\alpha}{\tilde{\chi}}
   {\partial}_{\beta}{\rho}-2{\hat{g}}^{{\alpha}{\beta}}{\partial}
   _{\alpha}{\phi}{\partial}_{\beta}e^{-2\phi}\right.\nonumber \\
& &\left.+4{\lambda}^2e^{2({\rho}-{\phi})}-\frac{\kappa}{4}
   {\hat{g}}^{{\alpha}{\beta}}{\partial}_{\alpha}{\chi}{\partial}
   _{\beta}{\chi}-{\frac 1 2}\sum_{i=1}^{N}{\hat{g}}^{{\alpha}{\beta}}
   {\partial}_{\alpha}f_{i}{\partial}_{\beta}f_{i}\right\}\nonumber \\
& &-\int d{\Sigma}\sqrt{{\pm}h}K\tilde{\chi}-\int d^{2}{\sigma}{\partial}
   _{\alpha}\left[\sqrt{-\hat{g}}{\hat{g}}^{{\alpha}{\beta}}\tilde{\chi}
   {\partial}_{\beta}{\rho}\right]
\end{eqnarray}
where $\hat{R}$ is the curvature scalar for ${\hat{g}}_{{\alpha}{\beta}}$, and
for simplicity, the factor $(2{\pi})^{-1}$ in front of action (2) has been
omitted (later it will be recovered). The last term in (5) is a surface term
coming from the relation $\sqrt{-g}R=\sqrt{-\hat{g}}\hat{R}-2{\partial}
_{\alpha}(\sqrt{-\hat{g}}{\hat{g}}^{{\alpha}{\beta}}{\partial}_{\beta}{\rho})$.

   In order to obtain the RST Hamiltonian explicitly, we need a field
redefinition to diagonalize the kinetic term of action (5), which is first
given by
\begin{eqnarray}
{\psi}_0&=&\frac{1}{\sqrt{\kappa}}e^{-2\phi}-\frac{\sqrt{\kappa}}{2}\phi
           +\sqrt{\kappa}\rho \nonumber \\
{\psi}_1&=&-\frac{\sqrt{\kappa}}{2}\chi +\sqrt{\kappa}\rho \nonumber \\
{\psi}_2&=&\frac{1}{\sqrt{\kappa}}e^{-2\phi}+\frac{\sqrt{\kappa}}{2}\phi
\end{eqnarray}
where ${{\psi}_{1}}$ is the hidden dynamical field [10-12], which was omitted
in the previous semiclassical consideration.

  From (6), the action (5) can be written as
\begin{eqnarray}
S&=&\int d^{2}{\sigma}\left\{\frac{\sqrt{\kappa}}{\omega}(\dot{{\psi}_0}
    -\dot{{\psi}_1}){\theta}^{\prime}+\frac{\sqrt{\kappa}}{\omega}
    ({\psi}_0^{\prime}-{\psi}_1^{\prime})({\omega}{\omega}^{\prime}
    -{\theta}{\theta}^{\prime})\right.\nonumber \\
 & &\left.+{\frac 1 2}{\omega}{\hat{g}}^{{\alpha}{\beta}}{\partial}_{\alpha}
    {\psi}_{\mu}{\partial}_{\beta}{\psi}_{\nu}{\eta}^{{\mu}{\nu}}
    +2{\lambda}^2{\omega}e^{\frac{2}{\sqrt{\kappa}}({\psi}_0-{\psi}_2)}
    -{\frac 1 4}{\omega}\sum_{i=1}^{N}{\hat{g}}^{{\alpha}{\beta}}
    {\partial}_{\alpha}f_{i}{\partial}_{\beta}f_{i}\right\}\nonumber \\
 & &-\left\{\int d{\Sigma}\sqrt{{\pm}h}K\tilde{\chi}+\int d^{2}{\sigma}
    {\partial}_{\alpha}\left[\sqrt{-\hat{g}}{\hat{g}}^{{\alpha}{\beta}}
    \tilde{\chi}{\partial}_{\beta}{\rho}\right]\right.\nonumber \\
 & &\left.\left.\left.+\int d{\tau}\left[\frac{\tilde{\chi}}{\omega}({\omega}
    {\omega}^{\prime}-{\theta}{\theta}^{\prime})\right]\right|^
    {{\sigma}=+\infty}
   _{{\sigma}=-\infty}+\int d{\sigma}\left(\frac{\tilde{\chi}}{\omega}
   {\theta}^{\prime}\right)\right|^{{\tau}=+\infty}_{{\tau}=-\infty}\right\}
\end{eqnarray}
where the expression for ${\hat{R}}$ has been used, and ${{\mu}=0,1,2}$, with
${\eta}^{{\mu}{\nu}}=(1,-1,-1)$. In the above, dots and primes denote
differentiation with respect to time and space respectively. The canonical
momenta associated with the fields $\{\omega,\theta,{\psi}_{\mu},f_i\}$ are
\begin{eqnarray}
P_{\omega}&=&0 \\
P_{\theta}&=&0 \\
P_0 &=& -\frac{\dot{\psi}_0}{\omega}+\frac{{\theta}{\psi}^{\prime}_0}{\omega}
        +\frac{\sqrt{\kappa}{\theta}^{\prime}}{\omega} \\
P_1 &=& -\frac{\dot{\psi}_1}{\omega}-\frac{{\theta}{\psi}^{\prime}_1}{\omega}
        -\frac{\sqrt{\kappa}{\theta}^{\prime}}{\omega} \\
P_2 &=& \frac{\dot{\psi}_2}{\omega}-\frac{{\theta}{\psi}^{\prime}_2}{\omega} \\
{\pi}_{i} &=& \frac{f_{i}}{2\omega}-\frac{{\theta}f_{i}^{\prime}}{2\omega}
\end{eqnarray}
Clearly (8) and (9) are primary constraints and $\omega(x)$ and
$\theta(x)$ play the role of lagrange multipliers. Then action (7) becomes
\begin{eqnarray}
S&=&\int d{\tau}\int d{\sigma}\left[\dot{{\psi}_0}P_0+ \dot{{\psi}_1}P_1
    +\dot{{\psi}_2}P_2+\sum_{i=1}^{N}\dot{f_i}{\pi}_i-({\omega}{\cal H}
    _{\omega}+{\theta}{\cal H}_{\theta})\right] \nonumber \\
 & &-\left\{\int d{\Sigma}\sqrt{{\pm}h}K\tilde{\chi}+\int d^{2}{\sigma}
    {\partial}_{\alpha}\left[\sqrt{-\hat{g}}{\hat{g}}^{{\alpha}{\beta}}
    \tilde{\chi}{\partial}_{\beta}{\rho}\right]\right.\nonumber \\
 & &+\int d{\tau}\left[\frac{\tilde{\chi}}{\omega}({\omega}
   {\omega}^{\prime}-{\theta}{\theta}^{\prime})-\sqrt{\kappa}({\psi}_0^{\prime}
   -{\psi}_1^{\prime}){\omega}\right.\nonumber \\
 & &\left.\left.\left.\left.+\sqrt{\kappa}(P_0+P_1){\theta}\right]
    \right|^{{\sigma}=+\infty}_{{\sigma}=-\infty}
    +\int d{\sigma}\left(\frac{\tilde{\chi}}{\omega}{\theta}^{\prime}\right)
    \right|^{{\tau}=+\infty}_{{\tau}=-\infty}\right\}
\end{eqnarray}
where
\begin{eqnarray}
{\cal H}_{\omega}&=&-{\frac 1 2}(P^2_0+{{\psi}_0^{\prime}}^2)
                    +{\frac 1 2}(P^2_1+{{\psi}_1^{\prime}}^2)
                    +{\frac 1 2}(P^2_2+{{\psi}_2^{\prime}}^2) \nonumber \\
  & &+\sqrt{\kappa}({\psi}_0^{{\prime}{\prime}}-{\psi}_1^{{\prime}{\prime}})
     -2{\lambda}^2e^{\frac{2}{\sqrt{\kappa}}({\psi}_0-{\psi}_2)}
     +\sum_{i=1}^{N}({\pi}_i^2+{\frac 1 4}{f_i^{\prime}}^2)=0 \\
{\cal H}_{\theta}&=&P_0{\psi}_0^{\prime}+P_1{\psi}_1^{\prime}
                    +P_2{\psi}_2^{\prime}-\sqrt{\kappa}(P_0^{\prime}
                    +P_1^{\prime})+\sum_{i=1}^{N}{\pi}_if_i^{\prime}=0
\end{eqnarray}
are secondary constraints, which satisfy the following Poisson brackets:
\begin{eqnarray}
\{{\cal H}_{\omega}(\sigma),{\cal H}_{\omega}({\sigma}^{\prime})\}
 &=&[{\cal H}_{\omega}(\sigma)+{\cal H}_{\omega}({\sigma}^{\prime})]
    {\partial}_{\sigma}{\delta}({\sigma}-{\sigma}^{\prime}) \nonumber \\
\{{\cal H}_{\theta}(\sigma),{\cal H}_{\theta}({\sigma}^{\prime})\}
 &=&[{\cal H}_{\theta}(\sigma)+{\cal H}_{\theta}({\sigma}^{\prime})]
    {\partial}_{\sigma}{\delta}({\sigma}-{\sigma}^{\prime}) \nonumber \\
\{{\cal H}_{\omega}(\sigma),{\cal H}_{\theta}({\sigma}^{\prime})\}
 &=&[{\cal H}_{\omega}(\sigma)+{\cal H}_{\omega}({\sigma}^{\prime})]
    {\partial}_{\sigma}{\delta}({\sigma}-{\sigma}^{\prime})
\end{eqnarray}
In the conformal gauge (${{\omega}=1,{\theta}=0}$), ${g_{++}=g_{--}=0}$,
${g_{+-}=-{\frac 1 2}e^{2\rho}}$, the action (14) can be reduced to
\begin{eqnarray}
S&=&\int d{\tau}\int d{\sigma}\left[\dot{{\psi}_0}P_0+ \dot{{\psi}_1}P_1
    +\dot{{\psi}_2}P_2+\sum_{i=1}^{N}\dot{f_i}{\pi}_i-{\cal H}_{\omega}\right]
    \nonumber \\
 & &\left.-\int d{\tau}\left[2\tilde{\chi}{\rho}^{\prime}
    -{\tilde{\chi}}^{\prime}
    \right]\right|^{{\sigma}=+\infty}_{{\sigma}=-\infty}
\end{eqnarray}
In the derivation of action (18), we have exploited the induced metric and
extrinsic curvature on the spacelike boundary $\Sigma$ defined by
$h=e^{2\rho}$ and $K={\nabla}_{\alpha}n^{\alpha}$,
while the induced metric and
extrinsic curvature on the timelike boundary B defined by $h=-e^{2\rho}$
and $K={\nabla}_{\alpha}{\gamma}^{\alpha}$ [20-22], where the timelike unit
vector ${n^{\alpha}}$ normal to $\Sigma$ and spacelike unit vector
${{\gamma}^{\alpha}}$ normal to B are defined in the conformal gauge as
[20-22]
\begin{equation}
n^{\alpha}=(e^{-\rho},0), \qquad
{\gamma}^{\alpha}=(0,e^{-\rho})
\end{equation}
with
\begin{equation}
n{\cdot}n=-1, \qquad
{\gamma}{\cdot}{\gamma}=+1
\end{equation}

    From (18), the total Hamiltonian can be written as
\begin{equation}
H_T=\int d{\sigma}{\cal H}_{\omega}+H_{\alpha}
\end{equation}
where
\begin{equation}
H_{\alpha}=\left.\left(2\tilde{\chi}{\rho}^{\prime}
           -{\tilde{\chi}}^{\prime}\right)
           \right|^{{\sigma}=+\infty}_{{\sigma}=-\infty}
\end{equation}
Now since ${\cal H}_{\omega}=0$ (weakly) is a constraint of the theory, the
energy is entirely given by the boundary term. Then we have our expression for
the ADM mass:
\begin{equation}
E_{ADM}=\left.{\Delta}\left(\tilde{\chi}{\rho}^{\prime}
        -{\frac 1 2}{\tilde{\chi}}
        ^{\prime}\right)\right|^{\infty}_{-\infty}
\end{equation}
In the derivation of Eq.(23), we have defined
${\Delta}{\tilde{\chi}}={\tilde{\chi}}
 -{\tilde{\chi}}_{LDV}$, ${\Delta}{\rho}={\rho}-{\rho}_{LDV}$, where
${\tilde{\chi}}_{LDV}$, ${\rho}_{LDV}$ are the linear dilaton vacuum solution,
since we should measure energy relative to the  linear dilaton vacuum. And
from Eq.(22) to (23), one half factor has been recovered which was omitted
previously.

    In the asympotically Minkowski coordinates ${{\sigma}^{\pm}}$ which
connect with the Kruskal-like coordinates $x^{\pm}$ by
$e^{{\pm}{\lambda}{\sigma}^{\pm}}={\pm}{\lambda}x^{\pm}$, we have
${{\rho}={\phi}+{\lambda}{\sigma}}$, so Eq.(23) can be written as
\begin{equation}
E_{ADM}=\left.{\Delta}\left({\lambda}\tilde{\chi}
        +\tilde{\chi}{\phi}^{\prime}
        -{\frac 1 2}{\tilde{\chi}}^{\prime}\right)\right|^{\infty}_{-\infty}
\end{equation}
when dropping all quantum corrections, i.e., ${\tilde{\chi}=e^{-2\phi}}$,
then we have
\begin{equation}
E_{ADM}=\left.{\Delta}\left[e^{-2\phi}({\lambda}+2{\phi}^{\prime})\right]
        \right|^{\infty}_{-\infty}
\end{equation}
Eq.(25) shows that at classical level our mass formula can be reduced to that
of refs.[5,6].

    From (6), Eq.(24) can be written as
\begin{equation}
E_{ADM}=\left.{\Delta}\left[{\lambda}\sqrt{\kappa}({\psi}_0-{\psi}_1)
        +\sqrt{\kappa}
        ({\psi}_0-{\psi}_1){\phi}^{\prime}-\frac{\sqrt{\kappa}}{2}({\psi}_0
        ^{\prime}-{\psi}_1^{\prime})\right]\right|^{\infty}_{-\infty}
\end{equation}
In the Kruskal-like coordinates $x^{\pm}$, the hidden dynamical field
${\psi}_{1}$ can be chosen as [10-12]
\begin{equation}
{\psi}_1(x)=-\frac{\sqrt{\kappa}}{2}\ln(-{\lambda}^2x^{+}x^{-})
\end{equation}

   Since Eq.(26) contains the dilaton field $\phi$, as pointed out in
refs.[1,23], we have in the weak-coupling region ($e^{2\phi} \ll 1$), i.e.,
${\sigma}\rightarrow{\infty}$,
${\phi}_{2}{\simeq}\frac{1}{\sqrt{\kappa}}e^{-2{\phi}}$,
while in the strong-coupling region ($e^{2{\phi}}{\gg}{1}$), i.e.,
${\sigma}\rightarrow{-\infty}$,
${\psi}_{2} \simeq \frac{\sqrt{\kappa}}{2}{\phi}$, and we will apply this
approximation in the following calculations for the ADM and Bondi mass.

   Now let us evaluate the ADM mass for the static solution [9,23,24]:
\begin{eqnarray}
\sqrt{\kappa}({\psi}_0-{\psi}_1)&=&\frac{m_0}{\lambda}+\sqrt{\kappa}
                     ({\psi}_0-{\psi}_1)_{LDV} \nonumber \\
\sqrt{\kappa}{\psi}_2&=&\frac{m_0}{\lambda}+(\sqrt{\kappa}{\psi}_2)_{LDV}
\end{eqnarray}
where $m_0$ is a constant, and $({\psi}_{0}-{\psi}_{1})_{LDV}$ and
$({\psi}_{2})_{LDV}$ are the quantum solutions corresponding to the linear
dilaton vacuum of the classical theory which are given by
\begin{eqnarray}
\sqrt{\kappa}{({\psi}_0-{\psi}_1)}_{LDV}&=&
   e^{{\lambda}({\sigma}^{+}-{\sigma}^{-})}+\frac{\kappa}{4}
   {\lambda}({\sigma}^{+}-{\sigma}^{-}) \nonumber \\
\sqrt{\kappa}{({\psi}_2)}_{LDV}&=&
   e^{{\lambda}({\sigma}^{+}-{\sigma}^{-})}-\frac{\kappa}{4}
   {\lambda}({\sigma}^{+}-{\sigma}^{-})
\end{eqnarray}
where we have used Eq.(27) for ${\psi}_1$. From (26) and the approximation for
$\phi$ in the different regions, then the ADM mass in the absence of the RST
boundary for the static solution is
\begin{equation}
E_{ADM}=0
\end{equation}
which is consistent with the result in ref.[6].

  In the case of an incoming shock wave of {\em f} matter
$T^{f}_{++}=(ae^{{\lambda}{\sigma}^{+}_{0}}/{\lambda}){\delta}({\sigma}^{+}
-{\sigma}^{+}_{0})$, $T^{f}_{--}=0$ [1],
the conformal frame in which the solution is asymptotically Minkowski is
related to the $\sigma$ frame by ${\bar{\sigma}}^{+}={\sigma}^{+}$,
${\bar{\sigma}}^{-}=-(1/{\lambda})\ln(e^{-{\lambda}{\sigma}^{-}}
-\frac{a}{\lambda})$,
and the solutions are [9,22]
\begin{equation}
\sqrt{\kappa}({\psi}_0-{\psi}_1)=\left\{\begin{array}{ll}
\frac{m}{\lambda}+e^{{\lambda}({\bar{\sigma}}^{+}-{\bar{\sigma}}^{-})}
+\frac{\kappa}{4}\ln\left[e^{{\lambda}{\bar{\sigma}}^{+}}\left(
e^{-{\lambda}{\bar{\sigma}}^{-}}+\frac{a}{\lambda}\right)\right],
& {\bar{\sigma}}^{+} \geq {\bar{\sigma}}^{+}_{0}, \\
e^{{\lambda}{\bar{\sigma}}^{+}}\left(e^{-{\lambda}{\bar{\sigma}}^{-}}
+\frac{a}{\lambda}\right)
+\frac{\kappa}{4}\ln\left[e^{{\lambda}{\bar{\sigma}}^{+}}\left(
e^{-{\lambda}{\bar{\sigma}}^{-}}+\frac{a}{\lambda}\right)\right],
& {\bar{\sigma}}^{+} < {\bar{\sigma}}^{+}_{0}
\end{array}
\right.
\end{equation}
and
\begin{equation}
\sqrt{\kappa}{\psi}_2=\left\{\begin{array}{ll}
\frac{m}{\lambda}+e^{{\lambda}({\bar{\sigma}}^{+}-{\bar{\sigma}}^{-})}
-\frac{\kappa}{4}\ln\left[e^{{\lambda}{\bar{\sigma}}^{+}}\left(
e^{-{\lambda}{\bar{\sigma}}^{-}}+\frac{a}{\lambda}\right)\right],
& {\bar{\sigma}}^{+} \geq {\bar{\sigma}}^{+}_{0}, \\
e^{{\lambda}{\bar{\sigma}}^{+}}\left(e^{-{\lambda}{\bar{\sigma}}^{-}}
+\frac{a}{\lambda}\right)
-\frac{\kappa}{4}\ln\left[e^{{\lambda}{\bar{\sigma}}^{+}}\left(
e^{-{\lambda}{\bar{\sigma}}^{-}}+\frac{a}{\lambda}\right)\right],
& {\bar{\sigma}}^{+} < {\bar{\sigma}}^{+}_{0}
\end{array}
\right.
\end{equation}
where we have exploited Eq.(27) for ${\psi}_{1}$, and
$m=ae^{{\lambda}{\sigma}^{+}_{0}}$ is the mass of the classical black hole. In
evaluating the ADM mass, we need to compare with the LDV solution in the same
conformal frame, that is, (29) with $\sigma$ replaced by $\bar{\sigma}$. Then
the ADM mass without the RST boundary for the dynamical solutions (31) and (32)
is given by
\begin{equation}
E_{ADM}(\bar{\tau})=\left.m-\frac{{\kappa}{\lambda}^2}{8}(\bar{\tau}
+\bar{\sigma})
\right|_{\bar{\sigma} \rightarrow -\infty}-\frac{{\kappa}{\lambda}}{8}(\ln
\frac{a}{\lambda}+1)
\end{equation}
The above equation shows that the ADM mass is infinite at quantum level,
whereas it is well-defined at classical level $({\kappa}=0)$. The divergent
part comes from the fact that due to the quantum anomaly the solution in the
region ${\bar{\sigma}}^{+} < {\bar{\sigma}}_{0}^{+}$ does not go to the LDV as
$\bar{\sigma}\rightarrow -\infty$ once we insist that the solution for
${\bar{\sigma}}^{+} > {\bar{\sigma}}_{0}^{+}$ is asymptotically Minkowski.

  As we know, the Bondi mass describes the total energy minus the energy that
has been radiated away up to a given retarded time, so it should be evaluated
on a line which is asymptotic to ${\bar{\sigma}}^{-}=constant$ at
${\bar{\sigma}}^{+}\rightarrow{\infty}$ and to
${\bar{\sigma}}^{+}={\bar{\sigma}}_{1}^{+} < {\bar{\sigma}}_{0}^{+}$ on
${\bar{\sigma}}^{-}\rightarrow{\infty}$, then in analogy with the expression
(26) for the ADM mass, the Bondi mass can be defined as [6]
\begin{eqnarray}
E_{Bondi}&=&{\Delta}\left[{\lambda}\sqrt{\kappa}({\psi}_0-{\psi}_1)
+\sqrt{\kappa}({\psi}_0-{\psi}_1)({\partial}_{+}-{\partial}_{-}){\phi}
\right. \nonumber \\
& &\left.\left.-\frac{\sqrt{\kappa}}{2}
({\partial}_{+}-{\partial}_{-})({\psi}_0-{\psi}_1)
\right]\right|^{{\bar{\sigma}}^{+}=+\infty}_{{{\bar{\sigma}}^{-}=
-\infty},  {\bar{\sigma}}^{+}={\bar{\sigma}}^{+}_{1}}
\end{eqnarray}
{}From (31), (32)and (34), we have
\begin{equation}
E_{Bondi}=m-\frac{{\kappa}{\lambda}}{4}\ln\left(1+\frac{a}{\lambda}
e^{{\lambda}{\bar{\sigma}}^{-}}\right)-\frac{{\kappa}{\lambda}^2}{8}
{\bar{\sigma}}^{+}_{1}-\frac{{\kappa}{\lambda}}{8}(\ln\frac{a}{\lambda}+1)
\end{equation}
The above equation shows that due to the existence of the hidden dynamical
field ${\psi}_1$, the original divergent Bondi mass [6] becomes convergent and
unbounded from below, however, when the RST boundary is imposed, the Bondi
mass will have a  lower bound.

   Now we consider the situation with the RST boundary as done in ref.[9]. The
RST boundary conditions put ${\partial}_{\pm}{\psi}_{2}=f=0$
on the critical curve, which is
regarded as the left boundary of spacetime whenever it is timelike. The RST
boundary is at [9]
\begin{equation}
{\psi}_2={\psi}^{c}_{2}=\frac{1}{\sqrt{\kappa}}\left[\frac{\kappa}{4}
-\frac{\kappa}{4}{\ln}\frac{\kappa}{4}\right]
\end{equation}

   As pointed out in ref.[6], there is no solution to the boundary curve
equation for the static case (28) with $m_{0}>0$. This means that the boundary
trajectory stays behind the classical event horizon $(x^{-}=0)$ [25]. So the
left boundary of the spacetime can be set at the negative infinite space end
$\bar{\sigma}\rightarrow{-\infty}$. Then we have
\begin{equation}
E_{ADM}=0
\end{equation}

    For the collapse case with the RST boundary, the ADM mass is given by
\begin{equation}
E_{ADM}=m-\frac{{\kappa}{\lambda}}{4}(3+\ln\frac{\kappa}{4})\cdot
{\left[1+\sqrt{1+\frac{{\kappa}{\lambda}^2}{a^2}e^{-2{\lambda}\bar{\tau}}}
\right]}^{-1}
\end{equation}
where we have used the boundary trajectory equation (36)
$e^{{\lambda}{\bar{\sigma}}^{+}_{B}}(e^{-{\lambda}{\bar{\sigma}}^{-}_{B}}
+\frac{a}{\lambda})=\frac{\kappa}{4}$, i.e.,
$e^{{\lambda}{\bar{\sigma}}_{B}}=\frac{a}{\lambda}e^{{\lambda}\bar{\tau}}
\left[-1+\sqrt{1+\frac{{\kappa}{\lambda}^2}{a^2}e^{-2{\lambda}\bar{\tau}}}
\right]$
for the dynamical solution (32), while
$e^{{\lambda}({\bar{\sigma}}^{+}_{B}-{\bar{\sigma}}^{-}_{B})}=\frac{\kappa}
{4}$ for the LDV solution (29), and $({\psi}_0-{\psi}_1)^c=\frac{\kappa}{4}
+\frac{\kappa}{4}\ln\frac{\kappa}{4}$, $({\psi}_2)^c=\frac{\kappa}{4}
-\frac{\kappa}{4}\ln\frac{\kappa}{4}$ at the RST boundary.

   Eq.(38) shows that there is a new contribution to the ADM mass from the RST
boundary due to the existence of the hidden dynamical field ${\psi}_1$. When
${\bar{\tau}}\rightarrow{-\infty}$, the quantum ADM mass goes to the mass of
the collapsing matter.

   The corresponding Bondi mass with RST boundary conditions can be evaluated
from (34). The lower limit in (34) is replaced by a point on the critical
curve $({\psi}_{0}-{\psi}_{1})=({\psi}_{0}-{\psi}_{1})^{c}$,
${\psi}_{2}=({\psi}_{2})^{c}$ for ${\bar{\sigma}}^{+} <
{\bar{\sigma}}_{0}^{+}$.
At the upper end, however, there are two regions to consider. Calling the point
where the apparent horizon and the critical curve intersect
$({\bar{\sigma}}_{s}^{+}, {\bar{\sigma}}_{s}^{-})$, we have the region
${\bar{\sigma}}^{-} < {\bar{\sigma}}_{s}^{-}$ (region I of RST; see ref.[9] for
figure) and the region between the timelike boundary and
${\bar{\sigma}}^{-}={\bar{\sigma}}_{s}^{-}$ (region II of RST). In region II
the black hole has decayed and the solution is taken to be the LDV. In region
I the solution is the collapsing solution for
${\bar{\sigma}}^{+} > {\bar{\sigma}}_{0}^{+}$. Thus we have [9-12]
\begin{equation}
\sqrt{\kappa}({\psi}_0-{\psi}_1)=\left\{\begin{array}{ll}
\frac{m}{\lambda}+e^{{\lambda}({\bar{\sigma}}^{+}-{\bar{\sigma}}^{-})}
+\frac{\kappa}{4}\ln\left[e^{{\lambda}{\bar{\sigma}}^{+}}\left(
e^{-{\lambda}{\bar{\sigma}}^{-}}+\frac{a}{\lambda}\right)\right],
& {\bar{\sigma}}^{-} < {\bar{\sigma}}^{-}_{s}, \\
e^{{\lambda}({\bar{\sigma}}^{+}-{\bar{\sigma}}^{-})}
-\frac{\kappa}{4}\ln\left(e^{{\lambda}{\bar{\sigma}}^{+}}\cdot
e^{-{\lambda}{\bar{\sigma}}^{-}}\right)
+\frac{\kappa}{2}\ln\left[e^{{\lambda}{\bar{\sigma}}^{+}}\left(
e^{-{\lambda}{\bar{\sigma}}^{-}}+\frac{a}{\lambda}\right)\right],
& {\bar{\sigma}}^{-} \geq {\bar{\sigma}}^{-}_{s}
\end{array}
\right.
\end{equation}
and
\begin{equation}
\sqrt{\kappa}{\psi}_2=\left\{\begin{array}{ll}
\frac{m}{\lambda}+e^{{\lambda}({\bar{\sigma}}^{+}-{\bar{\sigma}}^{-})}
-\frac{\kappa}{4}\ln\left[e^{{\lambda}{\bar{\sigma}}^{+}}\left(
e^{-{\lambda}{\bar{\sigma}}^{-}}+\frac{a}{\lambda}\right)\right],
& {\bar{\sigma}}^{-} < {\bar{\sigma}}^{-}_{s}, \\
e^{{\lambda}({\bar{\sigma}}^{+}-{\bar{\sigma}}^{-})}
-\frac{\kappa}{4}\ln\left(e^{{\lambda}{\bar{\sigma}}^{+}}\cdot
e^{-{\lambda}{\bar{\sigma}}^{-}}\right),
& {\bar{\sigma}}^{-} \geq {\bar{\sigma}}^{-}_{s}
\end{array}
\right.
\end{equation}
{}From (34),(39) and (40), we have
\begin{equation}
E_{Bondi}=\left\{\begin{array}{lll}
m-\frac{{\kappa}{\lambda}}{4}\ln\left(1+\frac{a}{\lambda}
e^{{\lambda}{\bar{\sigma}}^{-}}\right)-\frac{{\kappa}{\lambda}}{8}
(3+\ln\frac{\kappa}{4}){\left[1+\frac{\lambda}{a}
e^{-{\lambda}{\bar{\sigma}}^{-}}\right]}^{-1},
& in & I \\
\frac{{\kappa}{\lambda}}{4}{\left(1+\frac{\lambda}{a}
e^{-{\lambda}{\bar{\sigma}}^{-}}\right)}^{-1}-\frac{{\kappa}{\lambda}}{8}
(3+\ln\frac{\kappa}{4}){\left(1+\frac{\lambda}{a}
e^{-{\lambda}{\bar{\sigma}}^{-}}\right)}^{-1},
& in & II
\end{array}
\right.
\end{equation}
Like the ADM mass, there is also a new contribution to the Bondi mass from the
boundary due to the hidden dynamical field ${\psi}_{1}$. It can be seen from
(41) that in the presence of the RST boundary, the Bondi mass has a lower
bound. For ${\bar{\sigma}}^{-}{\rightarrow}{-\infty}$,
$E_{Bondi}{\rightarrow}m$; while for
${\bar{\sigma}}^{-}{\rightarrow}{+\infty}$,
$E_{Bondi}{\rightarrow}-\frac{{\kappa}{\lambda}}{8}(1+\ln\frac{\kappa}{4})$,
that is, $E_{Bondi}$ turns to be
negative. This is not surprising, since even at
${\bar{\sigma}}^{-}={\bar{\sigma}}_{s}^{-}$, $E_{Bondi}$ has already turned to
be negative. On the other hand, the energy flow is discontinuous at
${\bar{\sigma}}^{-}={\bar{\sigma}}_{s}^{-}={\lambda}^{-1}\ln[\frac{\lambda}{a}
(e^{\frac{4m}{{\kappa}{\lambda}}}-1)]$:
\begin{equation}
E_{Bondi}({\bar{\sigma}}_{s}-0)-E_{Bondi}({\bar{\sigma}}_{s}+0)=
-\frac{{\kappa}{\lambda}}{4}\left(1-e^{-\frac{4m}{{\kappa}{\lambda}}}\right)
\end{equation}
This is just the effect of the thunderpop of RST and is caused by the fact
that when the collpase solution is matched to the LDV along the null line
${\bar{\sigma}}^{-}={\bar{\sigma}}_{s}^{-}$, the result is continuous but not
smooth. Thus Eq.(42) shows that our new mass formula derived from Hawking and
Horowitz's conjecture is quite reasonable.

    In summary, the ADM and Bondi mass for the RST model have been first
discussed from Hawking and Horowitz's argument. The boundary terms in the
Hamiltonian come directly from the boundary terms in the action, and do not
need to be added "by hand". The result shows that the previous expression for
the ADM and Bondi mass actually needs to be modified at the quantum level.
But our new mass formula can be reduced to that of refs.[5,6] at the classical
level. It has been shown that there is a new contribution to the ADM and
Bondi mass from the
hidden dynamical field ${\psi}_{1}$. In the absence of the RST boundary, the
ADM mass for the static solution is zero, while for the dynamical case with
collapsing matter, the ADM mass is infinite and positive, whereas the Bondi
mass is finite. On the other hand, in the presence of the RST boundary, it has
been found that the ADM mass for the static solutions with $m_{0}>0$ is zero,
while for the dynamical case, there is a new contribution to the ADM mass from
the RST boundary, which is just the consequence that the hidden dynamical
field modifies the total energy. However, the Bondi mass begins to turn into
negative at an intermediate time, and is discontinuous across a certain null
line  ${\bar{\sigma}}^{-}={\bar{\sigma}}_{s}^{-}$, that is, the thunderpop of
the RST model can also be reflected from our new Bondi mass formula.
\vskip 20mm
\noindent
{\bf Acknowledgement}
\par
  This work was supported in part by the European Union Under the Human
Capital and Mobility programme. J.-G. Z. thanks the Alexander von Humboldt
Foundation for financial support in the form of a research fellowship. Y.-G.
M. was supported by the Natural Science Foundation of Fujian Province under
the grant No.A95009.

\end{document}